\documentclass[prx,twocolumn,superscriptaddress,floatfix,nopacs,nofootinbib]{revtex4-2}
\usepackage{graphicx,amsfonts,amssymb,amsmath,hyperref,enumerate}

\usepackage{algorithm}
\usepackage{algpseudocode}
\usepackage{pgfplots}
\usepackage{pgf}
\usepackage{lmodern}
\usepackage{import}
\usepackage{xr}
\usepackage{enumitem}
\usepackage{soul}

\newif\ifhyper
\hypertrue
\ifhyper
\hypersetup{
   citecolor = {green},
   colorlinks = {true}, 
   urlcolor = {blue} 
}
\fi

\newcommand{\beq}{\begin{equation}}
\newcommand{\eeq}{\end{equation}}
\newcommand{\beqa}{\begin{eqnarray}}
\newcommand{\eeqa}{\end{eqnarray}}

\newcommand{\comment}[1]{}

\def\Longarrow{\protect\@lra}
\def\@lra{\relbar\joinrel\relbar\joinrel\relbar\joinrel%
          \relbar\joinrel\rightarrow}

\def\norm#1{\left\lVert#1\right\rVert}

\begin{document} 

\title{Multi-disk clutch optimization using quantum annealing}

\author{John D. Malcolm}
\affiliation{Multiverse Computing, Centre for Social Innovation, 192 Spadina Avenue Suite 509, Toronto, ON M5T 2C2 Canada}

\author{Alexander Roth}
\author{Mladjan Radic}
\affiliation{ZF Friedrichshafen AG,
Research and Development, Graf-von-Soden-Platz 1,
88046 Friedrichshafen,
Germany}

\author{Pablo Mart\'{i}n-Ramiro}
\author{Jon Oillarburu}
\author{Borja Aizpurua}
\author{Rom\'{a}n Or\'{u}s}
\affiliation{Multiverse Computing, Parque Cientifico y Tecnol\'{o}gico de Gipuzkuo, Paseo de Miram\'{o}n, 170 3$^{\,\circ}$ Planta, 20014 Donostia / San Sebasti\'{a}n, Spain}

\author{Samuel Mugel}
\affiliation{Multiverse Computing, Centre for Social Innovation, 192 Spadina Avenue Suite 509, Toronto, ON M5T 2C2 Canada}

\begin{abstract}
In this work, we apply a quantum optimization algorithm to solve a combinatorial problem with significant practical relevance occurring in clutch manufacturing. It is demonstrated how quantum optimization can play a role in real industrial applications in the manufacturing sector. Using the quantum annealer provided by D-Wave Systems, we analyze the performance of the quantum and quantum-classical hybrid solvers and compare them to deterministic- and random-algorithm classical benchmark solvers. The continued evolution of the quantum technology, indicating an expectation for even greater relevance in the future is discussed and the revolutionary potential it could have in the manufacturing sector is highlighted.
\end{abstract}

\maketitle

\section{Introduction}
\label{sec:intro}

Optimization problems are omnipresent in all areas of science and industry. In particular, combinatorial optimization problems, which consist of searching for the global minimum of an objective function over discrete variables within a very large space of possible solutions, appear in many practical applications in every industry, including sectors such as finance, logistics, and manufacturing. For these problems, the size of solution space usually grows exponentially with the number of variables. Although specialized algorithms can be used to find (often-approximate) solutions for specific use cases, most optimization problems are intractable for sufficiently-large systems.

In recent years, the state of quantum computing technology has advanced to become practically relevant \cite{Gyongyosi_2019, Orus_2019, Luckow_2021}. These are machines which exploit quantum behavior (special phenomena that can occur at small scales and low temperature) to perform certain calculations which cannot be simulated efficiently on traditional, or \emph{classical}, computers.  For example, D-Wave Systems, an early player in the field, provides access to a particular type of analog quantum computer, referred to as a quantum annealer \cite{Boothby_2021}.
Quantum annealers have already been used with great success to solve hard optimization problems in multiple industrial applications, such as large scale production \cite{vw_paint, Streif_2021, Luckow_2021} and research and development \cite{bmw-challenge, seat-heating-ai} in manufacturing, financial investment strategy \cite{Mugel_2022}, shipping logistics \cite{Salehi_2022, Feld_2019, Moylett_2017}, and mobility services \cite{Yarkoni_2020, traffic_signals}.
There is an abundance of excitement around the advances that quantum computing is expected to yield in many areas.  
In this work, we focus on a challenging application to a particular manufacturing quality-control problem which is difficult to solve classically and is among the first studies applying quantum computing to a manufacturing problem (along with, for example, \cite{vw_paint}).
The problem is used as a platform to demonstrate how quantum computing can provide an edge in current industrial applications and to show its potential in the future as the technology continues to advance to evermore-powerful generations.

\begin{figure}
	\centering
	\includegraphics[width=0.48\textwidth]{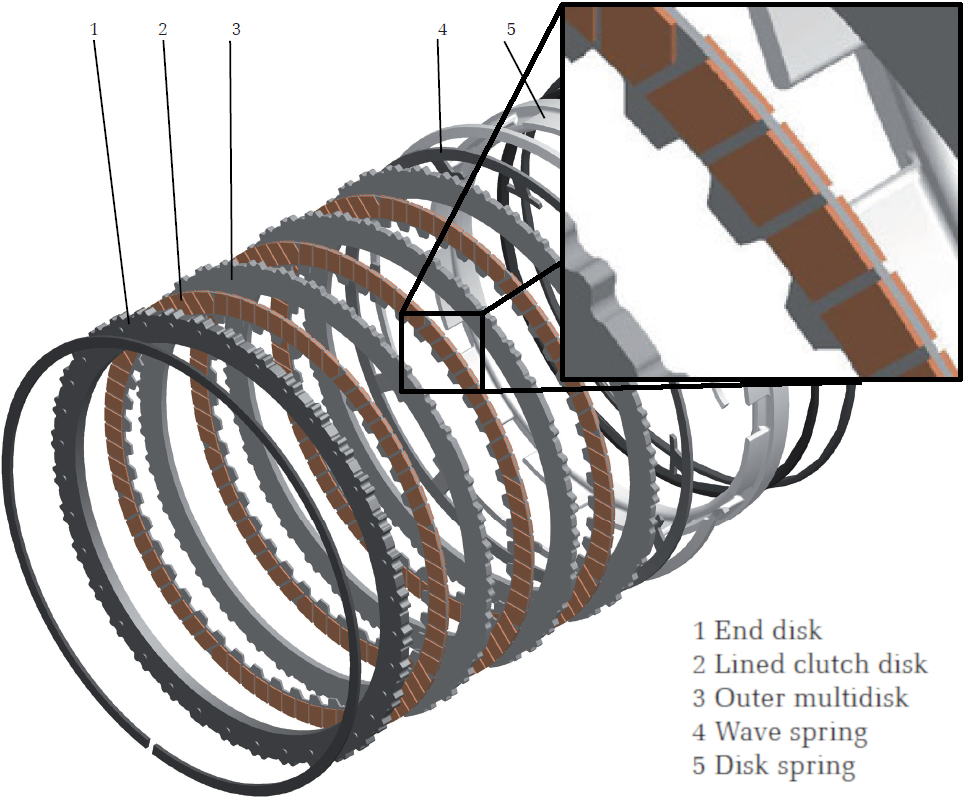}
	\caption{Disk sets in a multi-disk clutch contained in  the ZF 9HP transmissions. Disk set 2 (lined clutch disk) are the friction disks (f), which are padded with brown friction elements (see magnification). Disk set 3 (outer multidisk) are smooth metal disks (s).}\label{fig:clutch}
\end{figure}

In clutch manufacturing, one goal of shape or design optimization is to improve the friction of clutches and it plays a key role in large-scale production.
Previous works have considered such improvements by analyzing different designs and material combinations\cite{Thacker_2018, design_opt_clutch}.
The complementary approach in this work addresses the same goal, which we refer to as disk optimization, but with a different view and formulation of it.
More specifically, our approach to improve the friction in a multi-disk clutch consists of finding the optimal orientation for each of the disks.
For this purpose, we formulate the problem as a quadratic unconstrained binary optimization (QUBO) problem \cite{qubo_article} and introduce a quantum optimization algorithm to solve it using the D-Wave quantum annealer.
First, we show that the quantum solver is able to match the performance of current classical algorithms for small problems by using direct use of the QPU. 
Second, we demonstrate that a hybrid quantum-classical solver shows an exceptional performance for large-scale problems, greatly outperforming classical benchmarks. 
Therefore, our application of a quantum optimization algorithm provides an edge over classical solutions for large problem sizes when deployed on a hybrid quantum-classical solver.

In the following, Sec.~\ref{problem} provides details on the optimization problem and its mathematical formulation; Secs.~\ref{sec:methods} and~\ref{sec:approach} describe the techniques used in this study (quantum solvers and classical benchmarks) and the testing approach; results are presented and elaborated on in Secs.~\ref{sec:results} and~\ref{sec:discussion}; and we conclude the manuscript in Sec.~\ref{sec:conclusion}.

\section{Problem overview}
\label{problem}

In ZF Group, one optimization problem arises in the manufacturing of a multi-disk clutch (see Fig.~\ref{fig:clutch}). The multi-disk clutch contains two alternating sets of disks:
\begin{itemize}
	\item[(f)] Friction disks (metal disks padded with many friction elements); see brown paddings on \textit{lined clutch disk} (label 2) in Fig. \ref{fig:clutch}. 
	\item[(s)] Smooth metal disks; see \textit{outer multidisk} (label 3) in Fig. \ref{fig:clutch}.
\end{itemize}
Disks of the same kind are fixed to each other, while kinds (f) and (s) rotate freely relative to each other if the clutch is open. When the clutch closes, friction disks (f) start pressing  against the smooth disks (s) and assert a friction force reducing the relative rotation speed. If the clutch is closed the friction between disks in (f) and (s) is high enough that all disks have the same rotational speed.

The combinatorial optimization problem we are interested in arises due to manufacturing thickness variations of the disks (f) and the attached friction paddings. During manufacturing the friction disks (f) can each be stacked in $42$ different rotational positions, which are then locked in.  Thickness variations of each of the friction disks in a stack (e.g., 7 friction disks) add up and can lead to decreased performance of the clutch under load.  Such a decreased performance would be rejected in quality control tests after manufacturing.

The optimization question is therefore: In which of the $42$ possible rotations should each of the disks with friction paddings be placed to optimize the clutch performance and therefore maximize quality according to quality control?

Two of the relevant metrics to increase the clutch performance are:
\begin{enumerate}[label=(M{{\arabic*}})]
    \item \label{item:M1} The \emph{standard deviation} of the thickness deviations of the friction disks stack from the mean.
    \item \label{item:M2} The \emph{range} is defined as the difference between the highest and the lowest total thickness of the friction disks stack.
    \label{itm:M2}
\end{enumerate}
Both metrics are derived in Sec.~\ref{subsec:problem_formulation} and are given in Eqs.~\eqref{eqn:stdev} and~\eqref{eqn:range}, respectively.

\subsection{Problem Formulation}
\label{subsec:problem_formulation}

\begin{figure}
   \centering
   \includegraphics[width=0.48\textwidth]{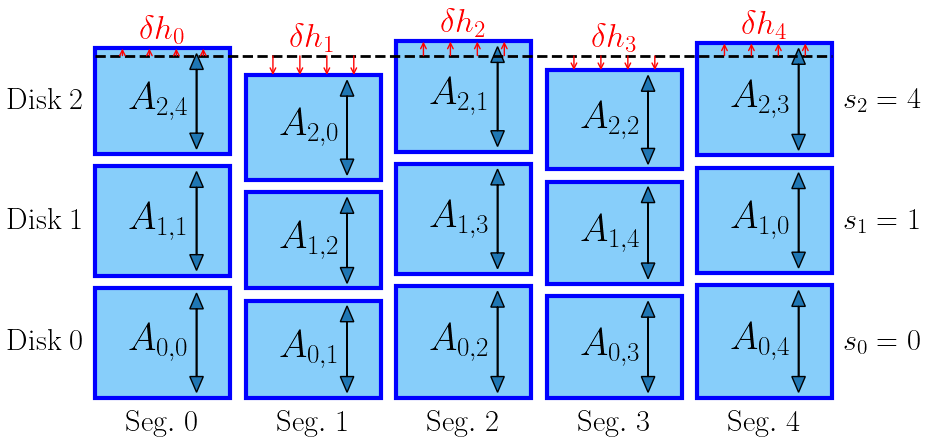}
   \caption{Illustration of the problem setup where the clutch is made up of a stack of rotatable disks (specifically the friction disks (f) in Fig.~\ref{fig:clutch}), with this example showing a 3-disk, 5-segment problem.  Disks are made up of discrete elements which line up with fixed segments.  The disks are labeled along the left and segments along the bottom, with disk shift numbers indicated on the right.  Element heights $A_{k,i}$ are labeled, where the indices correspond to the $k^{\rm th}$ disk and $i^{\rm th}$ segment in the stack's original configuration (prior to any shifts applied).  The horizontal dashed line gives the average segment height (number of disks, three, times the average element height) and the red arrows indicate each segment's deviation from the mean $\delta h_i$.}\label{fig:stack_terminology}
\end{figure}

The optimization problem focuses specifically on the clutch's set of friction disks (f) (Fig.~\ref{fig:clutch}), a stack of $N_D$ rotatable disks (Fig.~\ref{fig:stack_terminology}).  Each disk is made up of $N_S$ discrete elements (the friction pads) which line up within fixed segments used for reference.  As such, there are $(N_S)^{N_D}$ distinct configurations that the stack can take.  We denote the height of the $i^{\rm th}$ element of the $k^{\rm th}$ disk as $A_{k, i}$, with indexing starting at zero (hence, $i\in\{0, 1, \dots, N_S-1\}$ and $k\in\{0, 1, \dots, N_D-1\}$).  The height variation of an element is $B_{k, i} = A_{k, i} - \bar{A}$, where
\begin{equation}
   \bar{A} = \frac{1}{N_D N_S}\sum_{k=0}^{N_D-1}\sum_{i=0}^{N_S-1}A_{k, i}.
\end{equation}

The rotation of the $k^{\rm th}$ disk in the stack is captured by its shift number, $s_k\in\{0, 1, \dots, N_S-1\}$, or the number of discrete shifts to the left relative to its initial position.  Thus, following a rotation of the $k^{\rm th}$ disk, the height of the element shifted into the $i^{\rm th}$ fixed segment is $A_{k, i+s_k}$, where the summation in the index is understood to be modulo $N_S$.  Finally, this takes us to the $i^{\rm th}$ segment's height variation away from the mean segment height,
\begin{equation}
   \delta h_i = \sum_{k=0}^{N_D-1} B_{k, i+s_k}.
\end{equation}

The goal of this optimization problem is to find the combination of shift numbers which achieves as-close-as-possible-to uniformity in the full-stack height in line with the performance metrics standard deviation \ref{item:M1} and range \ref{item:M2}.  From the vector of segment height variations $\mathbf{\delta h} = (\delta h_0, \dots, \delta h_{N_S-1})$, these metrics \ref{item:M1} and \ref{item:M2}, respectively, are computed as
\begin{align}
\sigma(\mathbf{\delta h}) &= \sqrt{ \frac{1}{N_S} \sum_{i=0}^{N_S-1} \left(\delta h_i\right)^2 }, \label{eqn:stdev} \\
{\rm range}(\mathbf{\delta h}) &= \max(\mathbf{\delta h}) - \min(\mathbf{\delta h}) \label{eqn:range}.
\end{align}
Note that the mean of $\mathbf{\delta h}$ is zero and thus is not included in the definition of the standard deviation.

To form this as a programmable optimization problem, we can consider all segments in tandem, which yields
\begin{equation}\label{eqn:Ln_norm}
\min_{s_0, \dots, s_{N_D-1}} \, \norm{\mathbf{\delta h}}_n.
\end{equation}
That is, the optimization involves minimizing the $L_n$-norm of $\delta\mathbf{h}$ over the shift numbers.
For example, the $L_\infty$-norm minimizes the most extreme value in $\mathbf{\delta h}$ and is thus related to minimization of the range in segment height variations, though they are not equivalent in the sense that they do not lead to the same optimum.  To see the distinction, consider a vector $f$ with positive and negative elements $f_+$ and $f_-$, respectively.  Then,
\begin{align}
    \norm{f}_\infty &= \max \left( \norm{f_+}_\infty ,\, \norm{f_-}_\infty \right)
    ,
    \\
    {\rm range}(f) &= \norm{f_+}_\infty + \norm{f_-}_\infty
    .
\end{align}
While these are indeed different measures,  the range gives an upper bound on the $ L_\infty$-norm: ${\rm range}(f) \geq \norm{f}_\infty $.  Alternatively, the $L_2$-norm minimizes the vector elements' standard deviation, due to $\sigma(\mathbf{\delta h}) = \sqrt{1/N_S} \norm{\mathbf{\delta h}}_2$, and naturally lends itself to the formulation of the problem as a QUBO, as discussed in Sec.~\ref{sec:QUBO_formulation}.

\subsection{QUBO Formulation}
\label{sec:QUBO_formulation}

To make use of the D-Wave solvers (discussed in Sec.~\ref{sec:methods}), it is necessary to formulate the optimization problem as a QUBO \cite{qubo_article}.  That is, as an objective function with only linear or quadratic dependence on binary problem variables,
\begin{equation}
    \mathcal{H} = x^T Q x,
\end{equation}
where $x\in\{0,1\}^n$ is a vector of $n$ binary variables and $Q$ is an upper-triangular real-valued $n\times n$ matrix.  Note here that squaring a binary variable gives the same value of that variable ($x^2=x$), and hence any self-square terms in the expansion are considered to be linear.  From the general problem Eq.~\eqref{eqn:Ln_norm}, we minimize the square of the $L_2$-norm in order to fulfill the `quadratic' requirement, 
\begin{equation}\label{eqn:L2_norm}
    \norm{\mathbf{\delta h}}_2^2 = \sum_{i=0}^{N_S-1}\left(\delta h_i\right)^2.
\end{equation}

To map the shift numbers $(s_0, \dots, s_{N_D-1})$ onto binary variables, we use a unary encoding \cite{unary_encoding_article} (a.k.a., a `one-hot' encoding). That is, for each of the $N_D$ disks, we introduce a row vector $\mathbf{x}_k$ made up of $N_S$ binary variables $x_{k,j}\in\{0, 1\}$ (or an $N_D\times N_S$ matrix of variables) such that
\begin{eqnarray}\label{eqn:binary_encoding}
   x_{k, j} =
      \begin{cases}
         1 & {\rm if}\; j=s_k, \\
         0 & {\rm otherwise}.
      \end{cases}
\end{eqnarray}
For example, $\mathbf{x}_k=(1, 0, 0, \dots, 0)$ corresponds to $s_k=0$, $\mathbf{x}_k=(0, 1, 0, \dots, 0)$ corresponds to $s_k=1$, and so on.  This encoding allows us to take the problem variable $s_k$ from its role as an index (not applicable in a QUBO) and into usable function variables.  This is done by using the new $\{x_{k, j}\}$ variables as multiplicative assignment variables to indicate whether a particular disk element resides in a fixed segment or not.  In this way, the height variation of the $i^{\rm th}$ fixed segment can be written as
\begin{equation}\label{eqn:segment_height_variation}
   \delta h_i = \sum_{k=0}^{N_D-1}\sum_{j=0}^{N_S-1}B_{k,i+j}\,x_{k,j}.
\end{equation}

Using the unary encoding requires the introduction of $N_D$ constraints whereby one and only one of the variables in each vector $\mathbf{x}_k$ may be equal to one.  This is equivalent to requiring that the sum of the binary elements in $\mathbf{x}_k$ be equal to one.  This results in a linear equality constraint which can be added to the QUBO as a penalty term,
\begin{equation}\label{eqn:unary_encoding_constraints}
   \rho\left(\sum_{j=0}^{N_S-1}x_{k, j}-1\right)^2, \quad \forall k \in \{0, \dots, N_D-1\},
\end{equation}
where $\rho$ is a Lagrange multiplier, often referred to as the penalty strength \cite{penalty_term}.  Each of the above penalty terms (one for each disk) is equal to zero when the unary-encoding constraint is satisfied and is otherwise greater than zero.  Hence, adding these to the cost function with the appropriate penalty strength will force all infeasible solutions to have higher cost than the optimal feasible solution.  Ideally, one could set $\rho$ to be arbitrarily high, which works well, for example, when performing an exhaustive search over all possible solutions and ranking them by cost.
However, some of the solvers considered in this work (as discussed in Sec.~\ref{sec:methods}), have a maximum achievable resolution in the cost function.
That is, relative to the full range of all possible values achievable in the cost function, the solver can only distinguish between neighboring solutions if they are sufficiently separated from each other in cost.
As such, choosing the penalty strength too large can extend the range of the cost function to the point where it becomes difficult for the solver to resolve the optimal solution away from other feasible solutions.
In such applications, tuning the penalty strength is an important and non-trivial endeavor.

Combining Eqs.~\eqref{eqn:L2_norm}, \eqref{eqn:segment_height_variation}, and~\eqref{eqn:unary_encoding_constraints} yields the cost function,
\begin{equation}\label{eqn:free_cost_function}
   \begin{split}
   \mathcal{H} &= \sum_{i=0}^{N_S-1}\left(\sum_{k=0}^{N_D-1}\sum_{j=0}^{N_S-1}B_{k, i+j}\, x_{k,j}\right)^2 \\
   & \phantom{xxxxx} + \rho\sum_{k=0}^{N_D-1}\left(\sum_{j=0}^{N_S-1}x_{k, j}-1\right)^2, 
   \end{split}
\end{equation}
which has to be minimized.  This QUBO formulation requires $N_Q = N_D\cdot N_S$ binary variables and is fully connected.  That is, by expanding the cost function into a summation of linear and quadratic terms, at least one quadratic term exists for every possible pairing of problem variables.  As such, the problem amounts to the minimization of a fully connected graph problem where linear and quadratic coefficients correspond to vertex and edge weights, respectively (and any constant term corresponds to an overall offset which is irrelevant to the minimization).

One final adjustment can be made by exploiting a global rotational symmetry in the problem whereby rotating all disks by the same shift number leaves the cost function Eq.~\eqref{eqn:free_cost_function} invariant.  This operation is equivalent to changing the arbitrary reference point in the indexing (i.e., the choice of which segment is labeled as zero has no actual bearing on the physical clutch).  As such, we are free to leave the first disk frozen in its initial configuration (a shift number of zero), reducing the problem to finding the optimal rotation of all disks relative to the first.  Then, by setting $s_0=0$ or equivalently $\mathbf{x}_0=(1, 0, \dots, 0)$, Eq.~\eqref{eqn:free_cost_function} becomes
\begin{equation}\label{eqn:gauge_fixed_cost_fn}
   \begin{split}
      \mathcal{H} &= \sum_{i=0}^{N_S-1}\left(B_{0, i} + \sum_{k=1}^{N_D-1}\sum_{j=0}^{N_S-1}B_{k, i+j}\, x_{k,j}\right)^2 \\
      & \phantom{xxxxx} + \rho\sum_{k=1}^{N_D-1}\left(\sum_{j=0}^{N_S-1}x_{k, j}-1\right)^2.
   \end{split} 
\end{equation}
Exploiting the global rotational symmetry reduces the problem size to $N_Q = \left(N_D-1\right) \cdot N_S$ binary variables.

\section{Methods overview} 
\label{sec:methods}

We used a variety of methods and hardware implementations to address the disk optimization problem, with the goal of ascertaining the quantum solvers' performance against selected benchmarks on a classical computer.  These are listed below and each solver is detailed in the following.

\begin{enumerate}[label=\Alph*.]
    \item ZF Exact Classical Solver
    \item ZF Approximate Classical Solver
    \item D-Wave Simulated Thermal Annealer
    \item D-Wave Quantum Annealer
    \item D-Wave Leap Hybrid
\end{enumerate}

\subsection{ZF Exact Classical Solver}

This solver optimizes the range \ref{item:M2} and is \emph{guaranteed} to find one of the optimal configurations (there could be multiple) of the discrete shift numbers $(s_0, \dots, s_{N_D-1})$. It uses the `Branch and Bound' method \cite{Dakin1965ATA, DiscrProgramProbl}, consisting of two steps: the branch step followed by the bound step. The branch step divides the current problem under consideration in two or more subproblems. The bound step is responsible for estimating a lower (or upper) bound of the objective function with respect to an already divided subproblem. If the lower bound guarantees no optimal solution, then no further investigation of the whole branch is necessary and the whole branch is cut. This method is thus also sometimes called `Branch and Cut'. Therefore, the performance of this method stands and falls with the feasibility and `sharpness' of the bound. Note that, in the worst case, the complexity of the branch and bound method is $\mathcal{O}\big(N_S^{N_D-1}\big)$, identical to an exhaustive search, but an optimal solution is guaranteed.

\subsection{ZF Approximate Classical Solver}

As a result of the scaling mentioned above, large-scale problems are hard or even impossible to solve with the ZF Exact Classical Solver in a reasonable time.  To tackle this challenge, an approximate classical solver is proposed, subdividing the large scale case into smaller problems, which may be optimized efficiently. In short, the $N_D$ disks are split into $K$ subsets, where $K-1$ subsets contain $\lfloor N_D /K \rfloor$ disks and one subset contains the remaining $N_D - (K-1)\lfloor N_D /K \rfloor$ disks.  $K$ is chosen such that $\lfloor N_D/K \rfloor \leq 8$ and $K\geq3$. The entire subsets are optimized separately and then rotated relative to each other to find the optimum of the range \ref{item:M2}. Note, that this approach does not guarantee to find the global optimum because it only searches through a fraction of the whole solution space.

\subsection{D-Wave Simulated Annealer}
\label{subsec:sim_annealing}

Simulated thermal annealing (typically referred to as `simulated annealing') is a random algorithm which seeks to find the global minimum of an optimization problem while avoiding getting trapped in local minima.  By making use of a parameter analogous to temperature, the system is `hot' in the beginning to allow it jump to other states in successive iterations, so that it can escape local minima by hill climbing, and progressively `cooled' to converge to a local minimum (think of a popcorn kernel jumping around a hot kettle with many divots which is gradually cooled).  An advantage of this approach is that it allows the system to escape local minima when the temperature is sufficiently high.  However, it can struggle when there are many thin but deep minima, meaning that this random algorithm is not guaranteed to converge to the optimal solution.  Further details on this method can be found, for example, in \cite{simulated_annealing}.

In this work, we made use of the D-Wave simulated annealer available in the \texttt{dimod} python package \cite{dimod}.  This implementation allows for the user to input a QUBO, like the one in Eq.~\eqref{eqn:gauge_fixed_cost_fn}. The algorithm scales quadratically in the number of variables, and linearly in the number of \emph{sweeps} (steps of the algorithm) which corresponds to the number of different temperatures that are traversed. The initial state is selected at random, so that many runs of the algorithms can allow for a better exploration of the possible solutions.

\subsection{D-Wave Quantum Annealer}
\label{sec:dwave_qpu}

Quantum annealing is named in analogy to thermal annealing in that the physical process uses random quantum fluctuations (as opposed to thermal fluctuations).  The algorithm seeks out the optimal solution as the system is slowly transitioned such that its energy landscape (determined by the \emph{Hamiltonian}) reflects the cost function to be optimized.  By leveraging quantum behavior, the system begins in the lowest-energy configuration of a `mixing Hamiltonian' \cite{dwave_annealer_docs} which places the quantum bits (or \emph{qubits}) in a complete \emph{superposition} of all possible bitstring combinations at once.  By slowly transitioning to the `problem Hamiltonian', the quantum adiabatic theorem demonstrates that the quantum system will remain in the lowest-energy configuration, ultimately arriving at that which corresponds to the optimal solution of the cost function of interest \cite{adiabatic_quantum_computation}. Unwanted thermal fluctuations in the hardware can introduce noise to the system so that the optimal solution is not always obtained.  Hence, many samples are typically made in order to increase the likelihood of finding that optimum.

In this work, we made use of the D-Wave Advantage quantum processing unit (QPU), introduced in 2020. Here, over 5,000 physical qubits (superconducting loops held at the extremely low temperature of 15 mK) are laid out in a Pegasus graph architecture \cite{pegasus_graph} (where each qubit is generally connected to 15 other qubits in the system), so that it is possible to embed a fully connected graph with 180 vertices.\footnote{Minor embedding is the problem of replicating one graph onto another of different structure.  In this context, several physical qubits can be chained together with strong coupling to represent a single \emph{logical qubit}.} This is a major improvement compared to the previous generation D-Wave 2000Q processor, based on more than 2,000 qubits on a Chimera architecture \cite{chimera_graph} (connectivity per qubit of 6) where the largest fully connected graph embedding is of size 65.  In addition, the Advantage QPU is said to introduce less noise than earlier generation QPU's \cite{dwave_advantage}.  In the context of the QPU with reference to the minor graph embedding, we refer to a \emph{logical qubit} (not to be confused with an error-corrected qubit \cite{Egan_2021}) as a single binary variable appearing in the objective function.  A \emph{physical qubit} is a single superconducting loop in the QPU hardware.  Following the embedding, many physical qubits may be chained together to represent a single logical qubit.

\subsection{D-Wave Leap Hybrid}

The final solver we explored from D-Wave is their Leap Hybrid.  This uses a hybrid quantum-classical approach, whereby the submitted problem Eq.~\eqref{eqn:gauge_fixed_cost_fn} gets strategically decomposed into subproblems which are then passed in parallel to the D-Wave Advantage QPU described in Sec.~\ref{sec:dwave_qpu} as well as classical solvers \cite{dwave_annealer_docs}.  Following several iterations, Leap Hybrid then outputs the best solution obtained within a set run time.  As with the simulated annealer and the QPU, the Leap Hybrid is a probabilistic solver and is thus not guaranteed to produce the optimal solution, but has a high level of performance (as presented in Secs.~\ref{sec:results} and~\ref{sec:discussion} below).  For the moment, it can handle problem sizes of up to $10^6$ binary variables.

As a proprietary solver, the Leap Hybrid is understandably not entirely transparent.  This can make it difficult to compare against other solvers in some respects.  For example, a single run of this hybrid solver may make limited use of the QPU depending on the length of the queue for that quantum machine.  As such, it can be somewhat unclear exactly how much the quantum component contributed to finding a solution.  Nevertheless, we show below that the Leap Hybrid has exceptional performance for large problem sizes.  Alternative hybrid solvers are available which provide more transparency, such as the Kerberos solver available in the \texttt{dwave-hybrid} python package \cite{dwave_kerberos} (also developed by D-Wave, as the name suggests).  However, exploring alternative hybrid solvers was outside the scope of this project.

\section{Test Approach}
\label{sec:approach}

\subsection{Problem Instances}

To test the various methods, quantum and classical, we generate random matrices corresponding to $N_D$ disks with each $N_S$ elements.  We draw the element thickness $A_{k, i}$ out of a uniform distribution over the interval \mbox{$\left[ A_{0} - \frac \Delta 2, A_{0} + \frac \Delta 2 \right]$}, where $A_{0}$ is the target element thickness and $\Delta$ is the  maximal thickness variation of each element.

Relevant for the current manufacturing process in ZF are problem sizes with $N_D\leq 7$ and $N_S=42$ corresponding to $(N_D-1) \cdot N_S \leq 252 $ binary variables (or logical qubits) in the QUBO formulation.  Despite the practical limit, we explored problem sizes larger than this to analyze the performance and scaling of the different methods, giving us valuable insights in how to solve other fully connected QUBO problems in the future.

\subsection{Benchmark Testing}

Each of the random-algorithm solvers (Simulated Annealing, Quantum Annealing, and Hybrid) were run repeatedly a certain number of times in order to increase the probability of sampling the best solution.  These solvers, in contrast to the ZF solvers, aim to minimize the QUBO in Eq.~\eqref{eqn:gauge_fixed_cost_fn}, which corresponds to optimizing the standard deviation in Eq.~\eqref{eqn:stdev}.  The Simulated Annealing method acts as a random-algorithm benchmark for comparison against the quantum solvers, and the ZF solvers provide deterministic classical benchmarks.  In contrast to the QUBO formulation of the optimization problem, the ZF solvers focus on minimizing the range in Eq.~\eqref{eqn:range}.  We find, however, that an optimal solution with respect to standard deviation corresponds to a low (but not necessarily optimal) range, and vice versa.  In the following, we detail some of the specifications used in the three random-algorithm solvers.

\begin{itemize}
    \item Simulated Annealer - 35 samples were taken for each problem, using 1,500 as the number of sweeps.
    \item Quantum Annealer - From 20 up to 2,000 samples were taken for each problem, depending on the problem size, and annealing pauses were used to enhance the success probability as in \cite{power_of_pausing}.\footnote{Imposing an annealing pause involves adjusting the annealing schedule as the QPU transitions from the mixing Hamiltonian to the problem Hamiltonian.  As indicated in the cited work, a carefully positioned pause during the anneal can enhance the probability of finding the ground state of the problem Hamiltonian.}
    \item Hybrid Solver - 3 samples were taken for each problem.  Note, however, that in a single run of the hybrid solver, the backend runs iteratively over several subproblems and takes many samples directly from the quantum annealer in the process.
\end{itemize}

\section{Results}
\label{sec:results}

\par
\begin{figure*}
    \begin{center}
        \resizebox{\textwidth}{!}{\input{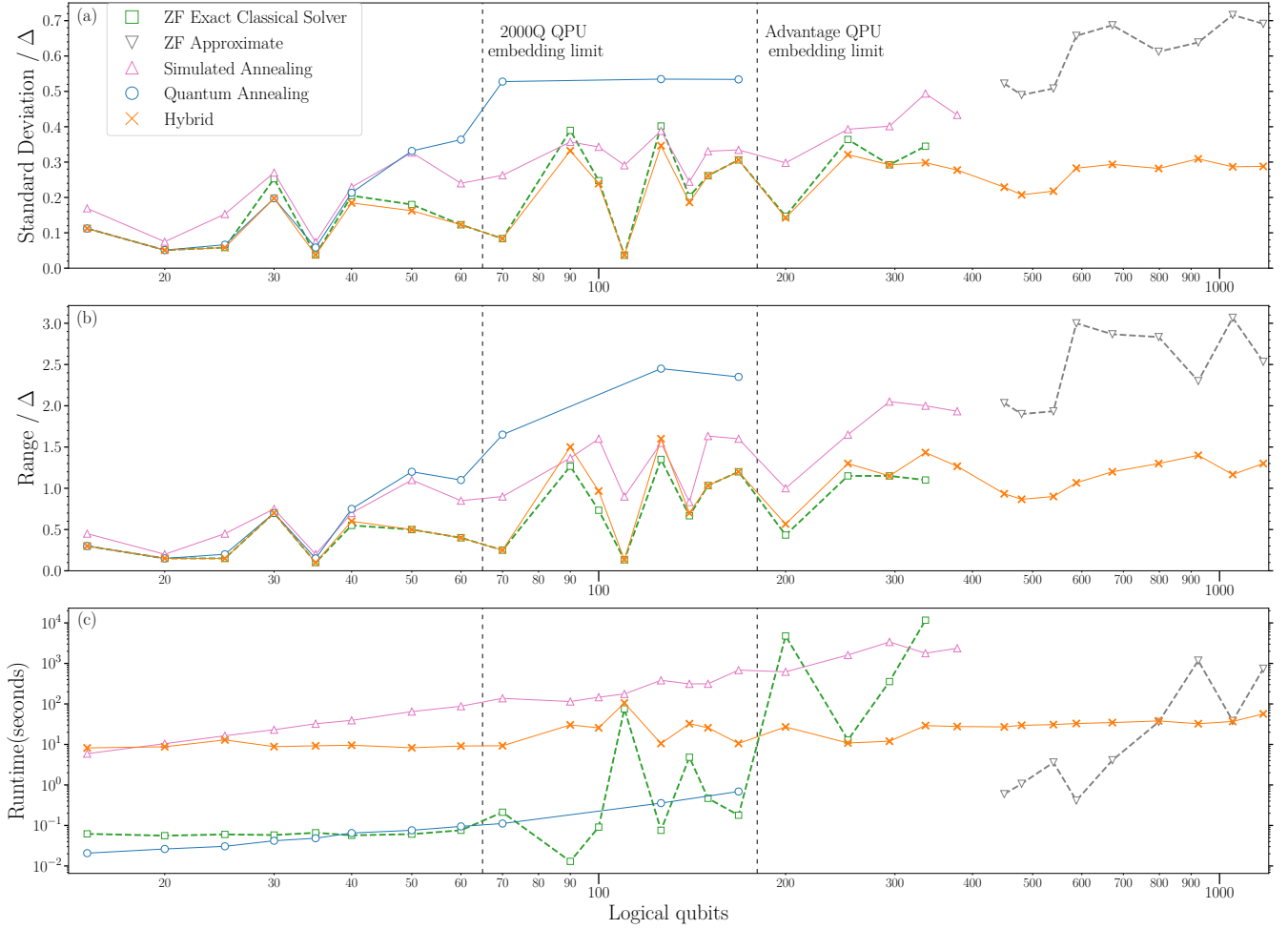}}
    \end{center}
    \caption{Comparison of (a) standard deviation (see Eq.~\eqref{eqn:stdev}), (b) range (see Eq.~\eqref{eqn:range}), and (c) runtimes for the different solvers across independent random disk optimization problem instances. For added clarity, solvers which optimized the standard deviation (range) are connected by solid (dashed) lines. Each problem corresponds to a fully connected graph with a certain number of logical qubits (e.g., the problem with $N_D=7$ disks and $N_S=42$ segments corresponds to $(N_D-1)N_S=252$ qubits).  Vertical dashed lines indicate the embedding limit of a fully connected graph in the Pegasus architecture of the D-Wave Advantage QPU on the right (used for both the Quantum Annealer and Hybrid solvers) and in the Chimera architecture of the D-Wave 2000Q QPU (indicated only for reference).  Note that the vertical scale in panel (c) and the horizontal scale in all three panels are logarithmic.}
    \label{fig:all_result_comparaison}
\end{figure*}

 In this section, we show the different results obtained by the considered solvers described in Sec.~\ref{sec:methods} under the approach described in Sec.~\ref{sec:approach}.  For varying problem sizes, the main quantities of interest are the problem metrics standard deviation, Eq.~\eqref{eqn:stdev}, and range, Eq.~\eqref{eqn:range}, associated with the solution that each solver outputs, as well as the runtimes.  All of these results are provided in panels (a), (b), and~(c) of Fig.~\ref{fig:all_result_comparaison}, respectively.  Note that all of the solutions presented in the figure were confirmed to satisfy the constraints in Eq.~\eqref{eqn:unary_encoding_constraints}.  For the Quantum Annealing solver, in particular, obtaining feasible solutions without sacrificing solution quality (i.e., low values for the metrics \ref{item:M1} and \ref{item:M2}) required careful tuning of the penalty strength $\rho$, which was carried out empirically.

\subsection{Problem Metric Comparisons}

To begin with, it is important to recall which quantities have been optimized by the different solvers. For the Simulated Annealing (pink triangles in Fig.~\ref{fig:all_result_comparaison}), Quantum Annealing (blue circles), and the Hybrid (orange crosses) solvers, the $L_2$-norm of $\mathbf{\delta h}$ is minimized (corresponding to minimization of the standard deviation, Eq.~\eqref{eqn:stdev}).  Data points for each of these solvers are connected by solid lines for added clarity. For the ZF Exact (green squares) and Approximate (gray inverted triangles) solvers, it is instead the range that is minimized, see Eq.~\eqref{eqn:range}, with data points in the figure connected by dashed lines.  Both of the quantities \ref{item:M1} and \ref{item:M2} are equally relevant and a solution is considered good or high quality if it is low in both (preferably optimal). The optimal solution with regard to the two different quantities can sometimes be different, but typically a configuration with minimal range yields a relatively low standard deviation, and vice versa. As a result, when comparing two solutions for the same problem instance, if one solver yields a lower standard deviation but a higher range than the other, we consider both solutions to be of a similar quality.

It is also important to note that the optimal metric value is instance dependent.  Any perceived trends with varying problem sizes (such as oscillations) are mostly coincidental.  However, the magnitude of the metrics \ref{item:M1} and \ref{itm:M2} tend to increase with the problem size.

In Fig.~\ref{fig:all_result_comparaison}(a), we see that for all problems considered, the Hybrid solver yields solutions with the lowest standard deviation compared to all other solvers.  While there is no guarantee from this solver that it finds the optimal solution, its consistently strong results across all investigated problems indicates that these solutions are at least close to optimal (especially for the smaller problem sizes where sampling errors are lower).  Direct use of the Quantum Annealing solver yields good solutions which are close to the Hybrid's solutions in standard deviation (though not necessarily optimal) up to problem sizes of 40 logical qubits.  Beyond that, sampling errors lead to results that are far above optimal, even for 50 qubits (well below the embedding limit of 180 qubits).  We further discuss these limitations of the Quantum Annealing solver in Sec.~\ref{sec:discussion}.  As for the benchmarks, the ZF Exact solver yields solutions that are close to optimal (if not optimal) in the standard deviation, despite its optimization of the range, see Eq.~\eqref{eqn:range}.  The high quality of solutions is consistent for all problem sizes up to its limit around 340-qubit equivalent (recall that the ZF solvers do not use the QUBO formulation of the problem and hence do not use the binary encoding of the problem variables in Eq.~\eqref{eqn:binary_encoding}).  Beyond this limit, the ZF Approximate solver is applied to large problems which yield solutions well separated from those obtained from the Hybrid solver.  Finally, the Simulated Annealing solver consistently yields suboptimal results (indicating the presence of deep local minima in the cost function).  In a practical setting, the results from the Simulated Annealing could be sufficient to pass quality-control requirements depending on the allowed tolerance, with the advantage that this solver is run locally.  However, it is also limited in that it cannot handle large problem sizes at about 400 qubits or greater, similar to the ZF Exact solver.  In such instances, the Hybrid solver is clearly preferred.

In Fig.~\ref{fig:all_result_comparaison}(b), the observations in the comparison of solutions' ranges are quite similar to the standard deviation results.  The main difference here is that the ZF Exact solver yields solutions with the guaranteed optimal range compared to the other solvers.  Interestingly, in many instances the Hybrid solver manages to find a solution with the same minimal value of the range, despite that solver's optimization of the standard deviation.  There are even instances, such as the one with 50 qubits, where the Hybrid solver matches the minimal range and still manages to find a lower standard deviation as compared to the ZF Exact solver.

\subsection{Runtime Comparisons}

Figure~\ref{fig:all_result_comparaison}(c) provides the total runtimes of each solver for the various problem sizes.  Note that comparisons of the absolute runtimes (i.e., the levels) are not fundamentally relevant given the different platforms each solver was run on, but are still useful for practical reasons.  For example, the Simulated Annealing solver was run locally on a MacBook Pro laptop, the ZF solvers on a ThinkPad, and the classical components of the Hybrid solver are run on supercomputing clusters.  More relevant is the relative scaling of each solver's runtime with the problem size that do not depend on the underlying hardware (beyond the classical vs. quantum qualities).  Furthermore, for the random-algorithm solvers (Simulated Annealing, Quantum Annealing, and Hybrid), a different, often more-relevant, measure is time to solution (which considers the probability of finding the optimal solution relative to a given level of confidence), rather than total runtime.  However, since the Simulated and Quantum Annealing solvers did not find the optimal solution in most instances we chose to instead present total runtime for a fixed number of samples.

For both the ZF Exact and ZF Approximate solvers, the runtime appears to be super exponential (increasing slope in the log-log plot), though the runtime of these approaches is very instance dependent.  For problem sizes corresponding to about 200 qubits or more (or even earlier, depending on usage requirements), the runtimes in the ZF Exact solver become prohibitive at the order of an hour or more.

At all problem sizes (small, medium, and large), the Simulated Annealing presents among the longest runtimes and scales quadratically (according to theory).  From a runtime perspective in practical applications, this approach becomes infeasible for medium-size problems and beyond.

Making direct use of the Quantum Annealing solver has very low runtimes, even for a problem size approaching its embedding limit.  For problem sizes below 40 qubits, where this solver was able to find close-to-optimal solutions (if not optimal), it had the fastest runtime compared to all other solvers (even when including internet latency).  This is particularly promising in that quantum computing technology is still rather nascent and solution quality is only expected to improve.  This topic is discussed further in Sec.~\ref{sec:discussion}.

The runtimes for the Hybrid solver are fairly consistent on the order of 10 seconds, based on the minimum fixed runtime (3 seconds) for each of the three samples made per problem.  Fluctuations seen in the runtime are due to internet latency and other time-costing overhead processes.  There is a shallow, almost imperceptible, increase in runtime (beyond 1,024 variables, the minimum runtime of the solver per sample gradually grows from the baseline 3 seconds).  For small- and medium-size problems, the ZF Exact solver is faster and yields similar-quality results as the Hybrid.  However, for large problem sizes the Hybrid solver is incredibly fast while still yielding high-quality results.  For a problem corresponding to 336 qubits, the ZF Exact solver took nearly 3.5 hours (11,659 seconds) while the Hybrid solved it in only 29 seconds.  The ZF Solver was unable to handle problems larger than this, while the Hybrid continued to yield high-quality results with a runtime of only 56 seconds on the largest problem considered (1,176 qubits).  The specifications indicate that it can handle problem sizes of up to $10^6$ variables.  In this work, the calculations were not pushed to these boundaries due to a lack of readily available benchmarks.

\section{Discussion}
\label{sec:discussion}

Based on the results presented in Sec.~\ref{sec:results}, the different scales of problems show differences in the performance of the solvers.  For small problems, below 50 qubits, all solvers find high-quality solutions except maybe the Simulated Annealing. However, in terms of runtime the Quantum Annealing solver is fastest---by several orders of magnitude than most solvers, and slightly better than the ZF Exact solver.  For small problem sizes, where sampling errors are reduced, taking 20 to 100 samples typically suffices to find (near-)optimal results, and so runtimes could be reduced without sacrificing solution quality as compared to larger problem sizes.

\begin{figure}
    \begin{center}
        \resizebox{.48\textwidth}{!}{\input{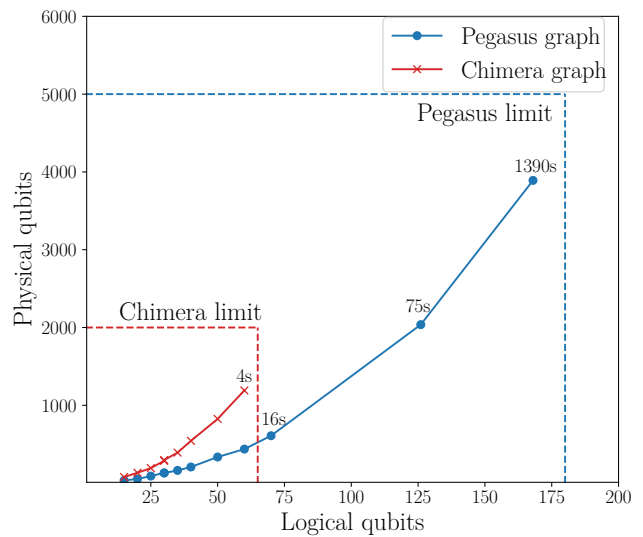}}
    \end{center}
    \caption{Number of physical qubits required for the embedding of fully connected graph problems on the Pegasus architecture (relevant to the Advantage QPU used in this study) and the Chimera architecture (in the previous-generation 2000Q QPU, included for reference).  The data points included here correspond to the problem sizes tested with the Quantum Annealing solver in Fig.~\ref{fig:all_result_comparaison}.  Numbers assigned to a subset of the data points indicate the runtime in seconds for calculating the minor embedding.}
    \label{fig:logical_qubit_vs_physical_qubit}
\end{figure}

For medium-size problems, 50 to 350 qubits, the Quantum Annealing solver no longer performs well.  This is due to noise in the machine arising from thermal fluctuations which lead to sampling errors.  In particular, as the number of problem variables increases linearly, embedding of the fully connected graph problem onto the Pegasus architecture (16-qubit connectivity) requires that evermore physical qubits be chained together to represent a single logical qubit.  This relationship is demonstrated in Fig.~\ref{fig:logical_qubit_vs_physical_qubit} which shows the number of physical qubits used for each of the problem sizes considered with this solver in blue (as well as the embedding limit assuming an availability of 5,000 physical qubits).  The physical qubit chain is achieved by imposing a strong ferromagnetic coupling across all qubits in the chain in an attempt to fix them all at the same value (either 0 or 1) which then corresponds to the value of the corresponding logical qubit.  However, for large problem sizes, chain breaks are more likely to occur (where the strong coupling fails to ensure all physical qubits in the chain are equal valued, potentially leading to an incorrect assignment in the corresponding logical qubit) because the chains are longer and are thus more susceptible to sampling errors.  In addition, there is more competition between the chains' couplings and the couplings which encode the optimization problem itself, obfuscating the details of the problem and making it more difficult for the solver to discern the optimal solution.

While there are clear limitations to the performance when making direct use of the Quantum Annealing solver, we point out again that this technology is still in its early stages and that there are major improvements with each new generation of quantum annealer from D-Wave.  To demonstrate this, Figs.~\ref{fig:all_result_comparaison} and~\ref{fig:logical_qubit_vs_physical_qubit} indicate the limits of embedding fully connected graphs onto the previous generation QPU, the 2000Q, compared to the Advantage QPU used in this study.  The latter figure shows that the increased connectivity in the Advantage QPU architecture means that fewer physical qubits are required to represent each logical qubit in the same size problem.  Had we made use of the 2000Q QPU, we would have only been able to run problem sizes up to about 65 qubits and the quality of the solutions would have been reduced relative to the Advantage QPU results.

As an aside, we also include the runtimes (in seconds) in Fig.~\ref{fig:logical_qubit_vs_physical_qubit} for determining the embedding of a subset of the problems considered (using the \texttt{minorminer} python package from D-Wave \cite{minorminer}).  Minor graph embedding is itself an NP-Hard problem and if done for each run of the Quantum Annealer solver would completely wash out the fast runtimes seen in Fig.~\ref{fig:all_result_comparaison}(c).  However, since all problems are known to be on fully connected graphs, we are able to precompute the embedding for each problem size and store this in a library for subsequent and repeated use.  As such, we did not include the embedding times in Fig.~\ref{fig:all_result_comparaison}(c).

It is in the latter half of this medium-size region (at about 200 qubits and greater) that the Hybrid solver becomes the preferred approach.  For all problem sizes it yields consistently high-quality results, and at this scale of problems it has a faster runtime than the ZF Exact solver.  Also in this region the runtimes of the ZF Exact solver increase rapidly, to a point where its use is prohibitively expensive in both runtime and memory constraints.  For large problem sizes, 350 qubits or more, the ZF Approximate solver has reasonable runtimes (up until about 800 qubits), but the solution quality is much poorer than in the Hybrid solver.

To rank the different solvers considered in this work, direct use of the Quantum Annealing solver is preferred for small problem sizes due to its capacity to find high-quality solutions in very short times. However, hardware limitations in that solver mean that both the Hybrid and ZF Exact solvers outperform it for medium-sized problems. In this region, these two algorithms yield similar results in terms of solution quality. In particular, we observe that the Hybrid solver finds solutions with slightly better standard deviation, whereas the ZF Exact solver finds solutions with a smaller range. For large-size problems, the Hybrid solver shows an impressive performance, clearly outperforming the ZF Approximate solver in terms of solution quality and runtime. In this region, the ZF Exact solver has no data points as it becomes prohibitively expensive in both runtime and memory, highlighting the benefits of the Hybrid solver.

\section{Conclusion and Outlook}
\label{sec:conclusion}

We have successfully applied a quantum optimization algorithm to tackle the optimization problem of improving the configuration of friction pads in a clutch, a challenging quality-control problem in manufacturing. As far as we are aware, this is among the first studies applying quantum computing to a manufacturing assembly problem (in league with, for example, \cite{vw_paint}). We presented the mathematical formulation of the problem in Sec.~\ref{subsec:problem_formulation}, a minimization of the range (for the deterministic classical benchmark algorithms) of height variations or the $L_2$-norm (for the random algorithms, including the quantum approaches that were the main focus of this work). For the latter, we derived a QUBO formulation using unary encoding for the shifts in order to meet the problem-formulation requirements for quantum annealing. We ran the algorithm on two quantum solvers (direct use of the Quantum Annealing solver and the quantum-classical Hybrid solver) on randomly generated problem instances of varying sizes and compared against classical benchmarks.  In a more general sense, this work is applicable to the optimization of fully connected QUBO problems with unary-encoding constraints.

From the results presented in Sec.~\ref{sec:results} and elaborated on in Sec.~\ref{sec:discussion}, we found that for small problem sizes (below about 50 qubits) the Quantum Annealing solver yields high-quality solutions (though not always optimal) with very short runtimes.  The performance of this solver is only expected to improve over time with further advances in the underlying technology.  For medium-size problems up to about 350 qubits, the Hybrid solver and the ZF Exact solver were on par in solution quality. However, in terms of runtime the latter rapidly scales upward as the number of variables is increased.  In contrast, the Hybrid solver remarkably has only limited scaling in runtime while maintaining high quality of solutions.  Using the Hybrid solver, we were able to solve large problems of up to 1200 fully connected variables, with relatively short runtime.  Furthermore, much larger problem sizes could have been run using the Hybrid solver, but without a useful benchmark for comparison we limited our investigations there.

With further advances in the quantum technology (extending, for example, the evolution of the D-Wave quantum annealer's capabilities highlighted in Fig.~\ref{fig:logical_qubit_vs_physical_qubit}), the benefits of making direct use of the QPU in combinatoric optimization is only expected to improve.
In addition to such improvements from technological advances, we believe that the current generation's performance could be made better with techniques available now that were outside the scope of this project.
For one, the penalty strength in Eq.~\eqref{eqn:gauge_fixed_cost_fn} could be optimized in a better way: an ideal penalty strength would limit the sampling of infeasible solutions picked by the solver, while preserving the details of the energy landscape of feasible solutions.  While we tuned the penalty strength empirically in this work, there are ideas available in the literature which suggest more systematic approaches (for example, in \cite{Ohzeki_2020, Huang_2021}).

From this work, we point out the strong performance of quantum-classical hybrid approaches, like in D-Wave's Leap Hybrid solver used here, which can lead to significantly better results than only classical or only quantum approaches. The results presented in this work demonstrate the competitive advantage that quantum computing can already offer in challenging problems encountered in the manufacturing sector. In our point of view, this study underlines the revolutionary potential of quantum computing in the manufacturing sector and the role it could play in the reshaping of industry.

In addition to our findings, future work could explore the use of domain-wall encoding to further enhance the efficiency of encoding optimization problems into physical qubits. Domain-wall encoding represents a variable with \(N\) possible states using \(N-1\) binary variables. This encoding method not only reduces the number of required qubits but also requires fewer and simpler penalty terms, which can improve the performance of quantum annealers. The potential benefits of domain-wall encoding have been demonstrated in various studies, showing improved embedding efficiency and performance in quantum optimization problems \cite{b35,b36,b37}. By incorporating domain-wall encoding, larger problem instances could be encoded more efficiently, potentially allowing quantum annealers to tackle even more complex and large-scale industrial problems.

This work is a collaboration between ZF Group and Multiverse Computing.  We would like to acknowledge and thank Amy Friske, Samuel Palmer, Serkan Sahin, and Devansh Trivedi for helpful discussions.

\bibliographystyle{apsrev4-2}
\bibliography{DiskOptimizationPaper}

\end{document}